\documentclass[a4paper]{quantumarticle}
\pdfoutput=1
\usepackage[utf8]{inputenc}
\usepackage[T1]{fontenc}
\usepackage{tikz}
\usepackage{graphicx}
\usepackage[numbers,sort&compress]{natbib}
\usepackage{bm}
\usepackage{physics}
\usepackage{mathtools}
\usepackage{textcomp}
\usepackage{amsmath}
\usepackage{amssymb}
\usepackage[colorlinks=true,citecolor=blue,linkcolor=blue,urlcolor=blue]{hyperref} 
 
\begin{document}
\date{September 10, 2019}
\title{Parallel in time dynamics with quantum annealers}
\author{Konrad Ja\l{}owiecki}
\affiliation{Institute of Physics, University of Silesia, Bankowa 12, 40-007 Katowice, Poland}  
\author{Andrzej Wi\k{e}ckowski}%
\affiliation{%
Department of Theoretical Physics, 
Faculty of Fundamental Problems of Technology,
Wrocław University of Science and Technology,  
50-370 Wrocław, Poland}
\author{Piotr Gawron}
\affiliation{Institute of Theoretical and Applied Informatics, Polish Academy of Sciences, Ba{\l}tycka 5, 44-100 Gliwice, Poland}
\author{Bart\l{}omiej Gardas}
\affiliation{Institute of Theoretical and Applied Informatics, Polish Academy of Sciences, Ba{\l}tycka 5, 44-100 Gliwice, Poland}
\affiliation{Institute of Physics, Jagiellonian University,  \L{}ojasiewicza 11, 30-348 Krak\'ow, Poland}

\begin{abstract}
Recent years have witnessed an unprecedented increase in experiments and hybrid
simulations involving quantum computers. In particular, quantum annealers. 
Although quantum supremacy has not been established thus far, there exist a
plethora of algorithms promising to outperform classical computers in the
near-term future.
Here, we propose a parallel in time approach to simulate dynamical systems
designed to be executed already on present-day quantum annealers. 
In essence, purely classical methods for solving dynamics systems are serial.
Therefore, their parallelization is substantially limited. In the presented
approach, however, the time evolution is rephrased as a ground--state search of
a classical Ising model. Such a problem is solved intrinsically in parallel by
quantum computers. The main idea is exemplified by simulating the Rabi
oscillations generated by a two-level quantum system (i.e. qubit)
experimentally.

\end{abstract}

\maketitle

\section{Introduction}%
It is needless to say that simulating dynamical systems with near-term quantum
technology poses one of the most difficult and technologically challenging
endeavor~\cite{feynman.60}. Various computations of certain aspects of many-body
quantum physics can already be assisted by the existing
hardware~\cite{king.carrasquilla.18,harris.sato.18,gardas.dziarmaga.18}. For
instance, recent experiments have demonstrated that quantum
annealers~\cite{kadowaki.nishimori.98} can be turned into neural networks that
can learn the ground state energy of a physical system~\cite{gardas.rams.18}. A
similar task can also be accomplished with fewer qubits using quantum
gates~\cite{kandala.mezzacapo.17,cervera-lierta.18}.

The aforementioned examples characterize static processes where there is no
real-time dynamics being simulated \emph{directly}. Noticeably, near-term
quantum annealers \emph{do} simulate quantum annealing, which is a
time-dependent phenomenon. However, the optimization problem itself, i.e., the
one to be solved by the annealer, exhibits no time dependence~\cite{lanting.14}.
Thus, following the time evolution, even of a single qubit on a quantum annealer
is a challenging task for the current technology. This should, nonetheless, be
possible at least in principle. Indeed, a time-dependent quantum problem can be
(re)formulated as a static one, defined on an appropriately enlarged Hilbert
space~\cite{mcclean.parkhill.13}. This is realized using the Feynman's clock
operator~\cite{caha.landau.18,biamonte.love.08}.

This observation naturally encapsulates a family of powerful algorithms referred
to as \emph{parallel in time} or \emph{parareal} methods, often invoked to
simulate the system's dynamics on heterogeneous classical
hardware~\cite{baffico.bernard.02,ruprecht.17}. The latter techniques
effectively take advantage of the fact that a part of the evolution can be
distributed and carried out in parallel. Nevertheless, with such an approach,
one can never reach full parallelism on any classical hardware (of the Turing
type) due to the communication bottlenecks~\cite{hill.marty.08}.

Nonetheless, these limitations do \emph{not} apply to the quantum hardware.
Quite the contrary, quantum computers operate in parallel and any algorithm (cf.
Refs.~\cite{shor.97,grover.97,harrow.hassidim.09}) they execute needs to be
carefully designed from scratch to utilize their intrinsic parallelism fully. 

As a proof of concept, in this article, we demonstrate that already the
present-day quantum annealers can be programmed to simulate dynamical systems in
parallel. In particular, we determine the time evolution of a single qubit (Rabi
oscillations) solely from experiments conducted on the newest D-Wave $2000$Q
quantum chip~\cite{rabi.37,kaluzny.goy.83,brune.schmidt.96,barnes.sarma.12}. At
the same time, due to the underlying connectivity (all-to-all) and the extensive
amount of qubits it requires, the proposed algorithm constitutes a natural test
which can determine the usefulness of various annealing technology realized by
e.g. the Floquet annealer~\cite{onodera.ng.19}, the large-scale
(photonic~\cite{pierangeli.marucci.19}) Ising
machines~\cite{mcmahon.marandi.16,inagaki.haribara.16,marandi.wang.14,inagaki.inaba.16},
and the Fujitsu digital annealer~\cite{aramon.rosenberg.19} in simulating
physical systems.

\section{Parallel in time dynamics}%
Consider a dynamical system (e.g. a quantum system isolated from its
environment~\cite{zurek.03}) whose behavior can be described by a $L$
dimensional and possibly time-dependent, Kamiltonian $K(t)$. The system dynamics
is encoded, at all times, in a (quantum) state, $\ket{\psi(t)}$, whose evolution
is governed by a Schr\"o{}dinger like equation~\cite{nielsen.chuang.10}, 
\begin{equation}
\label{eq:eq}
\frac{\partial \ket{\psi(t)}}{\partial t} = K(t)\ket{\psi(t)}.
\end{equation}
This first order differential equation admits a unique solution
$\ket{\psi(t)}:=U(t,t_0)\ket{\psi(t_0)}$, where
\begin{equation}
\label{eq:Ut}
U(t,t_0) = \mathcal{T} \exp\left(\int_{t_0}^{t} K(\tau) d\tau \right),
\end{equation}
propagates an arbitrary initial state, $\ket{\psi(t_0)}$, from $t_0$ to $t\ge
t_0$ whereas $\mathcal{T}$ denotes the time-ordering
operator~\cite{kosovtsov.04}. Such an ordering can be omitted whenever $[K(t),
K(t^{\prime})]=0$. In particular, for time independent systems, $\partial_t
K(t)=0$.
Furthermore, when $K(t)=-iH(t)/\hbar$ where $H(t)^{\dagger}=H(t)$ is a
Hamiltonian, the evolution operator~(\ref{eq:Ut}) is unitary and the
dynamics~(\ref{eq:eq}) is norm preserving and reversible, i.e.
$U(t,t^{\prime})^{\dagger}=U(t,t^{\prime})^{-1}=U(t^{\prime},t)$.

To solve Eq.~(\ref{eq:eq}), one usually discretizies the time interval $[t_0,t]$
selecting $N$ distinct moments, i.e. $t:=t_{N-1} > \dots > t_{n+1} > t_{n} >
\dots  > t_{0}$. The dynamics can then be formulated as a sequence of unitary
gates,
\begin{equation}
\label{eq:Ut2}
U(t,t_0) = U_{N-1} \cdots U_{n+1}U_{n} \cdots U_0,
\end{equation}
acting on an initial state. Note, each $U_n:=U(t_{n+1},t_n)$ can also be
formally expressed using Eq.~(\ref{eq:Ut}). Practically, however, for small time
steps, all gates $U_n$ are approximated using variety of
methods~\cite{nielsen.chuang.10}. Those include exact diagonalization for small
systems~\cite{iskakov.danilov.18}, Suzuki--Trotter
decomposition~\cite{hatano.suzuki.05}, commutator-free
expansion~\cite{alvermann.fehske.11} or sophisticated tensor networks
techniques~\cite{schollwock.05}.

The latter equation provides a starting point for various sequential numerical
schemes for solving differential equations on classical
computers~\cite{wanner.hairer.96}. In principle, however, those gates could also
be realized on a quantum computer, which could then resolve the unitary dynamics
efficiently~\cite{nielsen.chuang.10}. Unfortunately, current quantum hardware
does not allow for such gates to be constructed yet. Nevertheless, the
underlying idea behind decomposition~(\ref{eq:Ut2}) can be harnessed to
formulate an optimization problem that can be solved by present-day quantum
annealers~\cite{gardas.dziarmaga.18}. This is the main idea we put forward in
this work.

Indeed, consider a superposition of quantum states in different moments of time
$t_n$,
\begin{equation}
\label{eq:Psi}
\ket{\Psi} = \sum_{n=0}^{N-1} \ket{t_n} \otimes \ket{\psi(t_n)},
\end{equation}
where the \emph{clock} states are orthonormal, $\braket{t_n}{t_m}=\delta_{nm}$.
With the corresponding clock operator,
\begin{equation}
\label{eq:clock2}
 \mathcal{C} 
   = 
\sum_{n=0}^{N-2}
\big( 
\ket{t_{n+1}}\bra{t_{n+1}} \otimes I - \ket{t_{n+1}}\bra{t_n} \otimes U_n
+ \text{h.c.}
\big),
\end{equation}
one obtains $\mathcal{C} \ket{\Psi}=0\ket{\Psi}$. Thus, $\ket{\Psi}$ is the
ground state of $\mathcal{C}$. Obviously, this state is not unique since we have
specified neither initial nor boundary condition. However, introducing a
penalty, say $\mathcal{C}_{0}$, allows one to provide additional constrains. In
particular, specifying that
$\mathcal{C}_0=\ket{t_0}\bra{t_0}\otimes(I-\ket{\psi_0}\bra{\psi_0})$, the
following linear system
\begin{equation}
\label{eq:gsys}
\mathcal{A} \ket{\Psi} 
=
\ket{t_0}\otimes \ket{\psi_0},
\quad 
\mathcal{A}=\mathcal{C} + \ket{t_0}\bra{t_0} \otimes I,
\end{equation}
encodes Eq.~(\ref{eq:eq}) subjected to 
$\ket{\psi(t_0)}=\ket{\psi_0}$.
For hermitian systems, the above complex linear system of $N\times L$ equations
expresses the reversible dynamics of the system in terms of a sequence of
unitary gates~(\ref{eq:Ut2}). The hermitian clock operator can also be derived
from e.g. time-embedded discrete variational
principle~\cite{mcclean.parkhill.13}. The idea can be further extended to open
quantum systems~\cite{tempel.aspuru-guzik.14}. 

To solve the dynamics expressed in Eq.~(\ref{eq:gsys}) on a quantum annealer one
needs to formulate it as an optimization
problem~\cite{barahona.82,biamonte.love.08,lucas.14}. Moreover, such an
optimization needs to be encoded via the Ising spin-glass
Hamiltonian~\cite{kadowaki.nishimori.98} (or QUBO~\cite{wang.kleinberg.09})
defined on a particular \emph{sparse} graph called
chimera~\cite{dattani.chancellor.19} (or pegazus~\cite{dattani.szalay.19}).
Furthermore, at least complex fixed-point arithmetic is also required to express
quantum states in consecutive moments of time~\cite{chang.gambhir.18}. Here, we
incorporate a strategy introduced only recently in
Ref.~\cite{rogers.singleton.19}, cf. also Ref.~\cite{chang.gambhir.18} for real
matrices. To this end, we employ a natural correspondence between complex
numbers and real $2 \times 2$ matrices, namely
$a+bi \mapsto a \hat I+ib\hat \sigma_y$,
to represent $\mathcal{A}$ using only real entries.

We further rely on a straightforward observation that the solution to
Eq.~(\ref{eq:gsys}), expanded in the standard basis as $\ket{\bf x}=\sum
x_i\ket{i}$, also minimizes the following functional $h({\bf
x})=\|\mathcal{A}\ket{\bf x}-\ket{\Phi}\|^2$ and \emph{vice versa}. That is, a
global minimum of $h$, i.e. ${\bf x}_0$ is a solution~(\ref{eq:gsys}) as $h({\bf
x}_0)$=0. Moreover, when the simulated system is hermitian then $\mathcal{A}$ is
positive definite. Therefore, ${\bf x}_0$ is also a minimum of 
\begin{equation}
\label{eq:QUBO}
f({\bf x})=\frac{1}{2}\bra{\bf x}\mathcal{A}\ket{\bf x} -
\braket{\bf x}{\Phi}.
\end{equation}
as $\nabla f({\bf x}) = \mathcal{A}\ket{\bf x}-\ket{\Phi}$ and $\nabla^2 f({\bf
x})=\mathcal{A}>0$. Henceforward, we consider only hermitian systems and focus
exclusively on the latter equation~\footnote{This is mostly due to the technical
limitations (e.g. coupling's precision) of the current annealing technology.}. 

Since variables $x_i$ are real, the objective functions $f({\bf x})$ can
\emph{not} be programmed directly to be optimized on a quantum annealer.
Nevertheless, one can obtain the so called fixed-point representation for each
$x_i$ as a linear combination of \emph{binary} variables
$q_i^{\alpha}$~\cite{chang.gambhir.18}
\begin{equation}
\label{eq:float}
    x_i = 2^D \left(2 \sum_{\alpha=0}^{R-1}2^{-\alpha}q_i^{\alpha} -1\right).
\end{equation}
The above correspondence is constructed with the assumption that $R$ bits of
binary representation are used for every real number in the solution vector. In
our approach, the order of magnitude of the solution's coefficients is also
assumed, i.e. $x_i \in [-2^D, 2^D]$ for a fixed $D \in \mathbb{N}$. 

Therefore, the minimization problem to be solved on a quantum annealer can
finally be formulated as
\begin{equation}
\label{eq:QUBO2}
 f({\bf q}) = \sum_{i,\alpha} a_i^{\alpha} q_i^r + \sum_{i,j,\alpha,\beta} b_{ij}^{\alpha\beta} q_i^{\alpha} q_j^{\beta} + f_0,
\end{equation}
\def\A{\mathcal A}
where $b_{ij}^{\alpha\beta} = \A_{ij} 2^{1-\alpha-\beta+2D}$ and
\begin{eqnarray}
\begin{split}
a_i^\alpha &= \left( 2^{-\alpha+D}\A_{ii} - 2^D\sum_{j}\A_{ij}- \phi_i\right)2^{1-\alpha+D},
\\
f_0 &= 2^D\left( 2^{D-1}\sum_{ij}\A_{ij}+\sum_i \phi_i\right).
\end{split}
\label{eq:coeff}
\end{eqnarray}
The constant energy contribution, $f_0$, can be omitted as both $f({\bf q})$ and
$f({\bf q})-f_0$ have the same optimal solution ${\bf q}_0$. Since $f({\bf
q}_0)=f_0$, one can easily asses the quality of the solution found by any
heuristic approach.

\begin{figure}[t!]
\centering
    \includegraphics[width=\columnwidth]{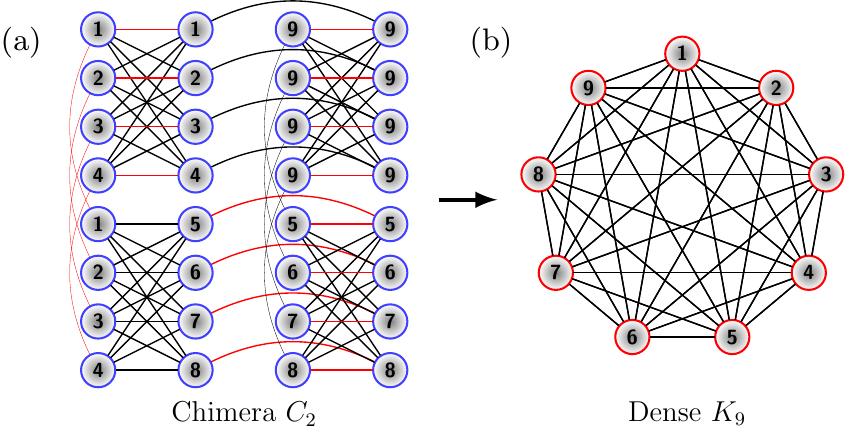}%
    \caption{   (a) An example of a sparse chimera graph [here $C_{2}$ (e.g.,
    $2\times 2\times 8$) consisting of $2 \cdot 2 \cdot 8=32$ qubits,
    cf.~Eq.~(\ref{eq:Hp})] and (b) the $9$ qubits complete graph $K_9$ embedded
    on $C_2$. Certain interactions on the chimera graph (marked as red)
    effectively ``glue'' physical qubits, $\hat \sigma^z_j$, to form logical
    variables, $q_i^{\alpha}$.}
    \label{fig:chimera}
\end{figure}

For small $N$, QUBO~(\ref{eq:QUBO2}) is defined on a complete graph [cf.
Fig.~\ref{fig:chimera}(b)] with $|\mathcal{V}|=R\times N \times (2L)$ vertices.
In contrast, when $N\gg 2$ the number of edges is equal to the number of nonzero
elements of $\mathcal{A}$ which is sparse. Currently, the biggest complete graph
that can be embedded on the $2000$Q chip has $|\mathcal{V}|=65$ vertices
($|\mathcal{V}|=180$ for the Pegazus topology~\cite{dattani.szalay.19}), cf.
Fig.~\ref{fig:chimera}. It is worth mentioning that classical solvers
(hardware-based or otherwise) usually offer better connectivity and thus can
realize much denser graphs without the need for embedding. For example, the
so-called coherent Ising machines (among others) can incorporate complete graphs
consisting of the order of $10^3$ vertices~\cite{inagaki.haribara.16}.
Therefore, QUBO generated from the dynamics provide a natural ``stress'' test
for those machines which can asses their usefulness in simulating physics.

\section{Quantum annealing}%
Adiabatic quantum computing can be seen as an alternative paradigm of
computation~\cite{kadowaki.nishimori.98}. Essentially, it is equivalent to the
gate model of quantum computation that uses logical gates operating on quantum
states to implement quantum algorithms~\cite{nielsen.chuang.10}. The main idea
is based on the quantum adiabatic theorem~\cite{avron.elgart.99}. When a system
starting from its ground state is driven slowly enough, it has time to adjust to
any change, and thus it can remain in the ground state during the entire
evolution.

Assume a quantum system is prepared in the ground state of an initial
(``simple'') Hamiltonian $\mathcal{H}_0$. Then, it will slowly evolve to the
ground state of the final (``complex'') Hamiltonian $\mathcal{H}_{\rm p}$ that
one can harness to encode the solution to an optimization problem. In
particular, the dynamics of the current D-Wave $2000$Q quantum annealer is
supposed to be governed by the following time-dependent Hamiltonian (cf.
Ref.~\cite{lanting.14})
\begin{equation}
\label{eq:dwave}
\mathcal{H}(s)/(2\pi\hbar)= -g(s) \sum_{i} \hat \sigma_{i}^x -\Delta(s) \mathcal{H}_{\rm p},
\quad
s \in [0, \tau],
\end{equation}
where the problem Hamiltonian $\mathcal{H}_{\rm p}$ realizes the spin-glass
Ising model defined on the chimera graph, $(\mathcal{E},\mathcal{V})$, specified
by its edges and vertices,
\begin{equation}
\label{eq:Hp}
\mathcal{H}_{\rm p} = \sum_{\langle i, j\rangle \in \mathcal{E}} J_{ij} \hat\sigma_i^z \hat\sigma_j^z + \sum_{i\in\mathcal{V}}h_i \hat\sigma_i^z.    
\end{equation}
\DeclareRobustCommand\tikzdot{\tikz[overlay,yshift=0.5ex] \fill[blue!90] (0.ex,0.ex) circle (0.3ex);}
\DeclareRobustCommand\tikzcircle{\tikz[overlay,yshift=0.5ex] \draw[red,thick] (0.ex,0.ex) circle (0.4ex);}
\DeclareRobustCommand\tikzquad{\tikz[overlay,yshift=0.ex,xshift=0.1ex] \draw[green,thick] (0.ex,0.ex) rectangle (1ex,1ex);}
\begin{figure}[t!]
    \centering
    \includegraphics[width=\columnwidth]{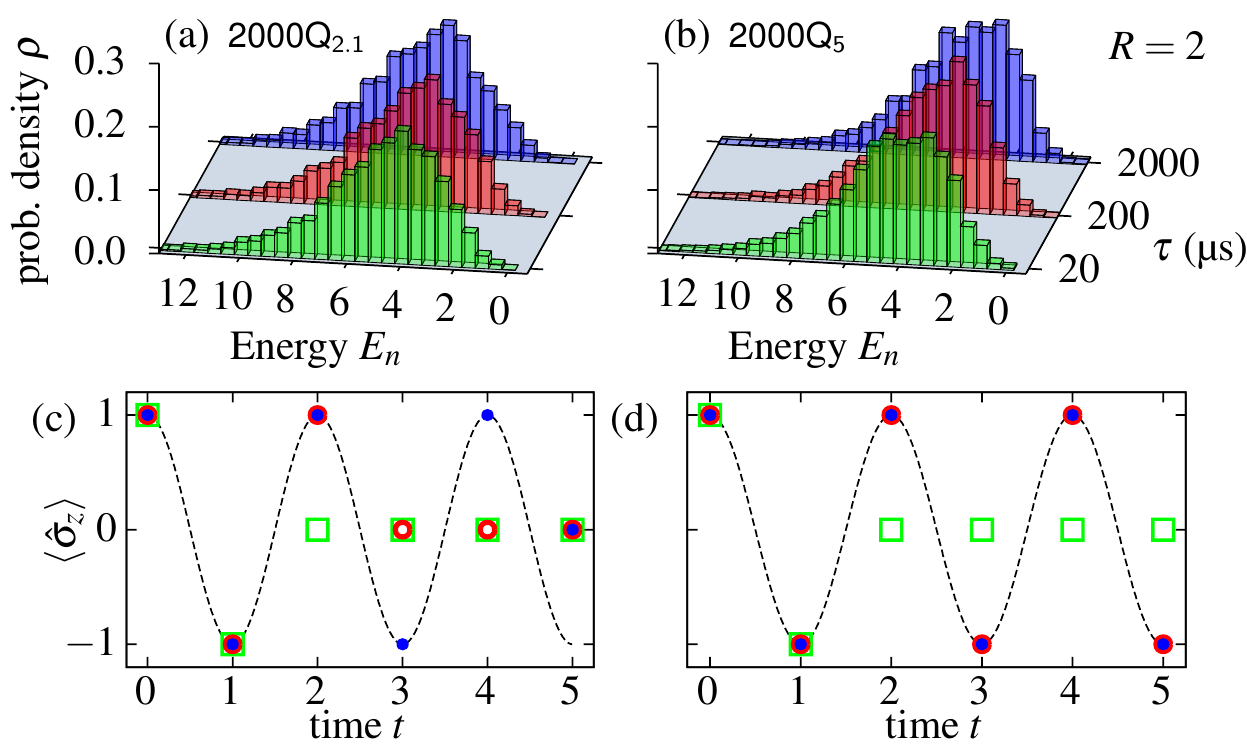}
    \caption{
     Rabi oscillations simulated on two generations of D-Wave quantum annealers.
     (a)--(b) the distribution of energy outputted by the annealers for
     different annealing times $\tau$. The two instances were generated from
     Eq.~(\ref{eq:QUBO2}) where $R=2$ bits of precision was assumed. The total
     number of variables in the corresponding QUBO was $|V|=168$. (c)--(d) the
     evolution in time of the spin $z$-component of a two level
     system~(\ref{eq:Sy}), $\omega=\pi/2$.
     (%
     \tikzquad\,\,\, -- 20 \textmu{}s,
     \,\tikzcircle\,\,\,-- 200 \textmu{}s,
     \,\tikzdot\,\,-- 2000 \textmu{}s%
     )
     }
    \label{fig:res2}
\end{figure}

The annealing time $\tau$ varies from microseconds to milliseconds depending on
the programmable schedule~\cite{lanting.14}. Typically, during the evolution
$g(s)$ varies from $g(0) \gg 0$ [i.e. all spins point in the $x$-direction] to
$g(\tau)\approx 0$ whereas $\Delta(s)$ is changed from $\Delta(0)\approx 0$ to
$\Delta(\tau) \gg 0$ [i.e. $\mathcal{H}(\tau) \sim \mathcal{H}_{\rm p}$]. Note,
the Hamiltonian $\mathcal{H}_{\rm p}$ is classical in a sense that all its terms
commute. Thus, its eigenstates translate directly to classical optimization
variables, $q_i^{\alpha}$, which we introduced to encode the time
evolution~(\ref{eq:gsys}) as QUBO~(\ref{eq:QUBO2}). The Pauli operators
$\hat\sigma_i^z$, $\hat\sigma_i^x$ describe the spin degrees of freedom in the
$z$- and $x$-direction respectively.

Dimensionless real couplers, $J_{ij}\in[-1,1]$, and magnetic fields, $h_i\in
[-2,2]$, are programmable. In practice, the actual values of those parameters
that are sent to the quantum processing unit differ from the ones specified by
the user by a small amount $\delta J_{ij}$, $\delta
h_i$~\cite{wieckowski.deffner.gardas.19}. This is due to various reasons
including noise effects which we will neglect in this work (cf.
Ref.~\cite{gardas.dziarmaga.18,gardas.deffner.18}).

Most practical optimization problems are defined on dense graphs which can be
embedded onto the chimera graph~\cite{choi.08}. There is, however, a substantial
overhead that effectively limits the size of problems that can be solved with
current quantum annealers. This is, nonetheless, an engineering issue that will
most likely be overcome in the near
future~\cite{dattani.szalay.19,onodera.ng.19}.

\begin{figure}[t!]
    \includegraphics[width=\columnwidth]{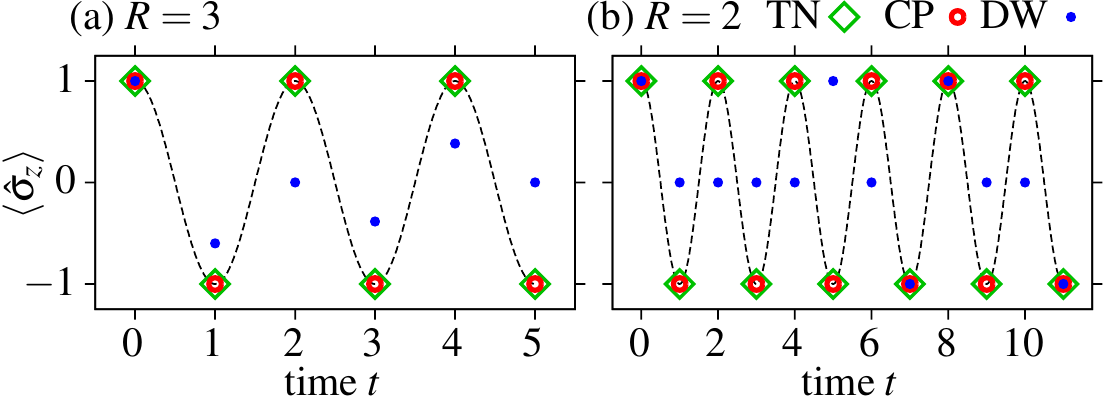}
    \caption{Performance of the two state of the art heuristic algorithms: the
    CPLEX optimizer (CP) and a recent solver based on tensor networks (TN) in
    comparison to the D-Wave $2000$Q quantum annealer (DW), cf.
    Fig~\ref{fig:res2}. The corresponding QUBO instances (encoded using double
    numerical precision) had total of $|V|=360$ and $|V|=624$ spin variables for
    (a), and (b) respectively. The annealing time was set to
    $\tau=200$\textmu{}s. The numerical precision of the solution vector is
    denoted as $R$.}
    \label{fig:res3}
\end{figure}

\begin{figure}[t!]
    \includegraphics[width=\columnwidth]{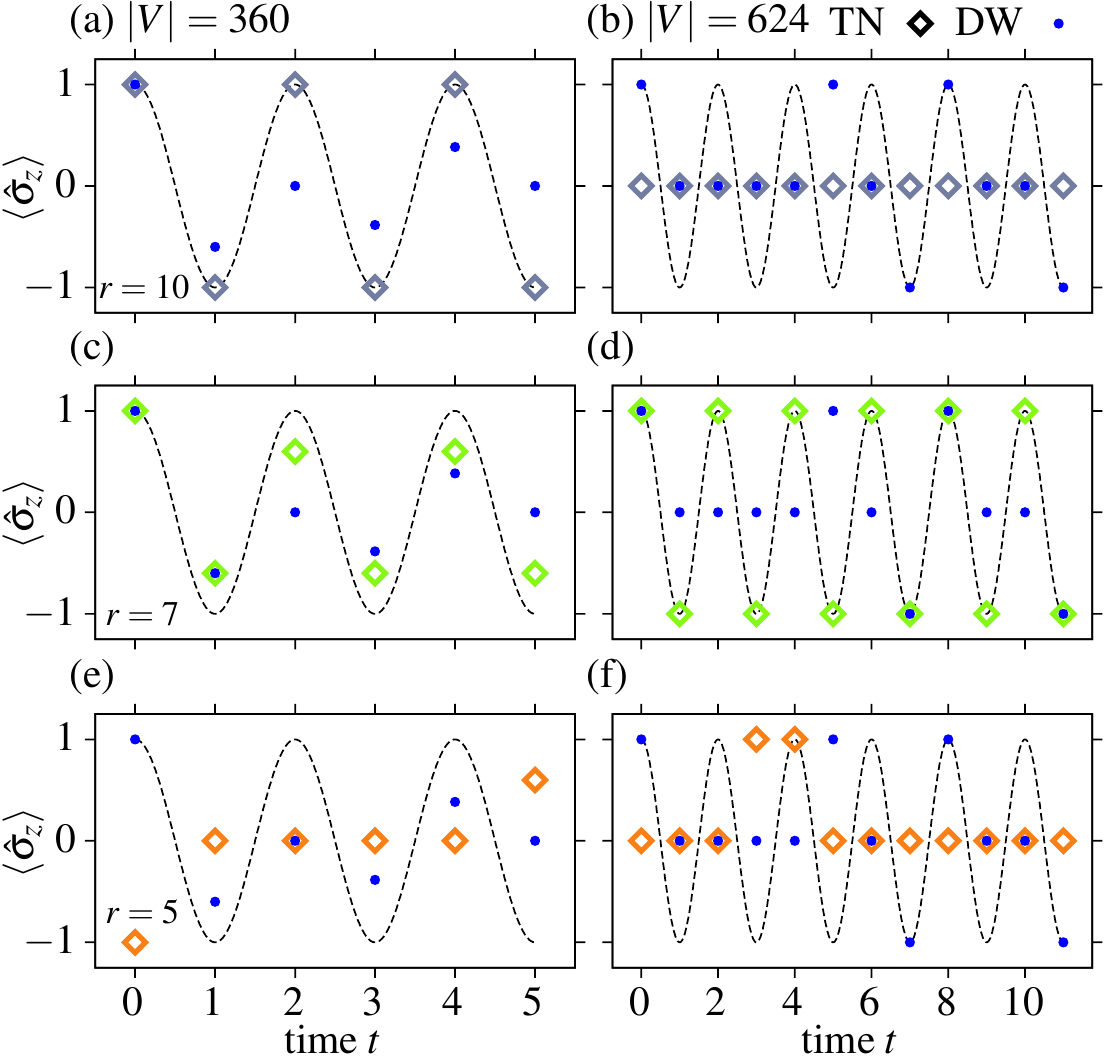}
    \caption{Degradation of the solution quality resulting from the truncation
    of the problem coefficients, cf. Eq.~(\ref{eq:coeff}), to a given numerical
    precision denoted as $r$. The numerical results were obtained by finding the
    ground state with tensor networks (TN). As a reference point, we included
    experimental data from the D-Wave $2000$Q quantum annealer (DW). This effect
    is expected to be predominant for the current quantum annealing technology.
    It is already visible on Fig.~\ref{fig:res2},~\ref{fig:res3} and it further
    increases with the increasing graph size $V$.}
    \label{fig:res4}
\end{figure}

\section{Results}%
To exemplify the main idea we consider a two-level quantum system (qubit) whose
Hamiltonian reads
\begin{equation}
\label{eq:Sy}
H = \omega \hat \sigma_y, 
\end{equation}
where $\hat \sigma_y$ is the Pauli spin matrix in the $y$-direction. For the
sake of simplicity, we further set $\omega=\pi/2$. 
Moreover, due to the limited number of qubits and sparse connectivity of D-Wave
quantum annealers, we mainly consider the system's evolution at six distinct
integer time points, starting from $\ket{\psi_0}=|0\rangle$. This ensures that
the dynamics can be captured precisely with two bits of precision per component
of the state vector, thus allowing one to run experiments on the D-Wave $2000$Q
annealer. For the illustrative purposes we reconstruct
$\langle\hat\sigma_z\rangle(t)$.

As depicted in Fig.~\ref{fig:res2}, the low noise D-Wave $2000$Q annealer was
able to capture the dynamics faithfully [cf. Figs.~\ref{fig:res2}(b), (d)], for
$\tau=200$~\textmu{}s, $2000$~\textmu{}s. This demonstrate an improvement in
comparison to the (not that) older generation, results for which are shown in
Figs.~\ref{fig:res2}(a), (c).

In contrast, results obtained from an emulation of the D-Wave output with tensor
networks (cf. Ref.~\cite{gardas.mohseni.rams.18}) are presented in
Fig.~\ref{fig:res3}. As a reference point, we have also included solutions found
by the CPLEX optimizer~\cite{cplex}. Both these solvers, being purely classical,
exhibit superior performance in comparison to the D-Wave quantum
annealers~\footnote{Assuming sufficiently large precision of all
$\mathcal{A}_{ij}$}. This is noticeable especially for problems that require
bigger graphs resulting from higher precision---($R\ge 3$, $N=6$), cf.
Fig.~\ref{fig:res3}(a)---or extra time points ($N>6$, $R=2$), cf.
Fig.~\ref{fig:res3}(b). Similar degradation of the solution quality with the
increasing problem size has been observed, e.g., in
Ref.~\cite{king.bernoudy.18,hamerly.ingaki.19} in the context of problems
requiring complete graphs, cf. Fig.~\ref{fig:chimera}. 

The behavior, as mentioned above, is expected from an early stage device which
is prone to errors. Their origins, however, are anything but straightforward to
pinpoint precisely. In stark contrast, there is yet another source of errors
that is related to the precision of $J_{ij}$, and $h_i$~\cite{dwave-ice}. Those
errors are believed to be predominant for the type of simulations introduced in
this work. Indeed, Fig.~\ref{fig:res4} shows the destructive (above all
\emph{not} monotonic) effect of the limited precision---$r$, of the problem
coefficients $\mathcal{A}_{ij}$---on the solution. Beyond a certain threshold,
neither the D-Wave annealer nor the aforementioned classical heuristics can
reproduce the dynamics (i.e., oscillations) accurately. 

\section{Conclusions}%
In this article, we have proposed a parallel in time approach to simulate
dynamical systems with the quantum annealing technology. Our results constitute,
first and foremost, a proof of concept demonstrating how the first generation of
quantum annealers can be employed to simulate the time evolution of simple (e.g.
two-level) quantum systems. This task is \emph{a priori} difficult for the
current prototypical quantum computers which has been designed mostly to
simulate static phenomena. 

Furthermore, not only the Ising instances we have generated can be executed on
the commercially available D-Wave annealers, but they can also be tested on:
coherent Ising
machines~\cite{pierangeli.marucci.19,mcmahon.marandi.16,inagaki.haribara.16,marandi.wang.14,inagaki.inaba.16},
the Floquet annealer~\cite{onodera.ng.19}, and the Fujitsu digital
annealer~\cite{aramon.rosenberg.19} that celebrate all-to-all connectivity. This
provides a practical ``stress'' test for those machines which can determine
their usefulness in simulating various time-dependent properties of physical
systems.

\begin{acknowledgments}
All authors are thankful to Marek M. Rams and Jacek Dziarmaga for very fruitful
discussions and comments. This work was supported by the National Science Centre
(NCN, Poland) under Grant No. 2015/19/B/ST2/02856 (KJ), 2016/22/E/ST6/00062
(PG), 2016/23/B/ST3/00647 (AW), 2016/20/S/ST2/00152 (BG) and NCN together with
European Union through QuantERA ERA--NET program 2017/25/Z/ST2/03028 (BG). BG
acknowledges the Google Faculty Research Award 2018. We gratefully acknowledge
the support of NVIDIA Corporation with the donation of the Titan V GPU used for
this research. 
\end{acknowledgments}

\bibliographystyle{apsrev4-1_nature}
\bibliography{ref} 

\end{document}